\documentclass[sigplan,screen,10pt]{acmart}

\AtBeginDocument{%
  \providecommand\BibTeX{{%
    \normalfont B\kern-0.5em{\scshape i\kern-0.25em b}\kern-0.8em\TeX}}}

\usepackage{graphicx}
\usepackage{hyperref}

\usepackage{subcaption}
\usepackage{changepage}
\usepackage{amsmath}
\usepackage{enumitem}

\newcommand{\pt}[1]{\textcolor{magenta}{#1}}
\newenvironment{myquery}{\medskip \begin{adjustwidth}{0.4cm}{}}{\end{adjustwidth} \medskip}
\newcommand{\mytext}[1]{\texttt{\small \fontfamily{cmtt}\selectfont #1}}

\setcopyright{rightsretained}
\acmPrice{}
\acmDOI{10.1145/3486607.3486750}
\acmYear{2021}
\copyrightyear{2021}
\acmSubmissionID{onward21papers-p10-p}
\acmISBN{978-1-4503-9110-8/21/10}
\acmConference[Onward! '21]{Proceedings of the 2021 ACM SIGPLAN International Symposium on New Ideas, New Paradigms, and Reflections on Programming and Software}{October 20--22, 2021}{Chicago, IL, USA}
\acmBooktitle{Proceedings of the 2021 ACM SIGPLAN International Symposium on New Ideas, New Paradigms, and Reflections on Programming and Software (Onward! '21), October 20--22, 2021, Chicago, IL, USA}

\begin{document}

\title{SkyQuery: An Aerial Drone Video Sensing Platform}

\author{Favyen Bastani}
\affiliation{%
  \institution{MIT CSAIL}
  \country{US}
}
\email{favyen@csail.mit.edu}

\author{Songtao He}
\affiliation{%
  \institution{MIT CSAIL}
  \country{US}
}
\email{songtao@csail.mit.edu}

\author{Ziwen Jiang}
\affiliation{%
  \institution{MIT CSAIL}
  \country{US}
}
\email{ziwenj@csail.mit.edu}

\author{Osbert Bastani}
\affiliation{%
  \institution{University of Pennsylvania}
  \country{US}
}
\email{obastani@seas.upenn.edu}

\author{Sam Madden}
\affiliation{%
  \institution{MIT CSAIL}
  \country{US}
}
\email{madden@csail.mit.edu}

\begin{abstract}
Video-based sensing from aerial drones, especially small multirotor drones, can provide rich data for numerous applications, including traffic analysis (computing traffic flow volumes), precision agriculture (periodically evaluating plant health), and wildlife population management (estimating population sizes). However, aerial drone video sensing applications must handle a surprisingly wide range of tasks: video frames must be aligned so that we can equate coordinates of objects that appear in different frames, video data must be analyzed to extract application-specific insights, and drone routes must be computed that maximize the value of newly captured video. To address these challenges, we built \emph{SkyQuery}, a novel aerial drone video sensing platform that provides an expressive, high-level programming language to make it straightforward for users to develop complex long-running sensing applications. SkyQuery combines novel methods for fast video frame alignment and detection of small objects in top-down aerial drone video to efficiently execute applications with diverse video analysis workflows and data distributions, thereby allowing application developers to focus on the unique qualities of their particular application rather than general video processing, data analysis, and drone routing tasks. We conduct diverse case studies using SkyQuery in parking monitoring, pedestrian activity mapping, and traffic hazard detection scenarios to demonstrate the generalizability and effectiveness of our system.
\end{abstract}



\maketitle

\begin{sloppypar}
\section{Introduction}

When equipped with sensors, computation, and communication capabilities, small multirotor aerial drones are an ideal mobile sensing platform for many applications. Indeed, video sensing on aerial drones has been proposed for a number of tasks, including {\it earth science}, for monitoring coastal sea ice and habitat loss~\cite{ryan2015}; {\it disaster emergency response}, for tracking floods, fires, and people in search-and-rescue~\cite{search-and-rescue}; {\it automated mapping}, for measuring traffic incidents, congestion, and parking~\cite{traffic}; {\it civil infrastructure monitoring}, for finding defects in bridges, buildings, and pipelines~\cite{bridges}; and {\it precision agriculture}, for micro-climate monitoring of plant growth~\cite{precision-ag}.

However, realizing applications that involve extracting insights from aerial drone video remains difficult; in fact, every one of the above applications required the authors to design an entire application-specific system from scratch, which is both costly and error-prone. These applications face a number of challenges. First, the unstructured nature of video means that machine learning models are needed to extract structured information, such as object detections, that can be further processed and analyzed. General-purpose models such as YOLOv3~\cite{yolov3} and Mask R-CNN~\cite{maskrcnn} that are standard in other computer vision tasks are not well-suited for detecting many object types in aerial drone video. For example, we find that these models deliver poor performance when applied to detect pedestrians and cyclists in drone video, even when fine-tuned on hand-labeled examples, because of the small footprint that these objects have in top-down images. Second, analysis on the raw outputs of machine learning models applied on aerial drone video is difficult since these outputs are represented in a pixel coordinate system corresponding to the camera frame that depends on the drone's position. For most applications, in order to reason about information such as object detections, pixel coordinates need to be transformed to world coordinates (longitude-latitude).

In this paper, we describe SkyQuery, a platform we built for rapidly developing and deploying long-running video sensing applications on aerial drones.
SkyQuery provides an expressive domain-specific language in which programs specify sensing-analytics-routing loops that define not only how insights should be extracted from video, but also where video should be captured in the future: a program applies a series of data processing operations to extract useful insights from video captured by drones, and then further processes these insights to specify how future drone routes should be computed for collecting additional video.
SkyQuery transparently performs efficient video frame alignment and robust object detection so that program developers can focus on application-specific analysis steps.
SkyQuery provides a library of analytics operators to encode these steps, including drone scheduling operators that define how SkyQuery should decide where drones should fly next: for example, a parking availability mapping application that tracks available parking spots across a city may require drones to fly more frequently over retail parking lots where parking exhibits unpredictable spikes than over residential lots; this requirement can be programmed in SkyQuery by leveraging its forecasting-based scheduling operator, which forecasts future observations (e.g., number of free spots) as a probability distribution, and prioritizes collecting video in areas with high-variance predictions. SkyQuery also incoporates a web interface to enable users to efficiently experiment with different workflows when developing new programs, and an aerial drone controller to automate takeoff, waypoint-following, landing, and recharging at a base of operations.

We develop novel methods for video frame alignment and object detection in aerial drone video to address the aforementioned challenges. Prior work in frame alignment is either too inaccurate or too computation-intensive for aerial drone video sensing, where we may need to simultaneously process several video streams. We propose an alignment method that leverages both sensor readings (GPS and compass) and image features to efficiently compute an accurate mapping from pixels in each video frame to world coordinates.
We also propose a change-aware object detection method specialized for detecting small, moving objects like pedestrians and cyclists in aerial drone video by comparing the image at the current frame with images captured earlier.

We evaluate SkyQuery both in simulation and in a real-world scenario consisting of eight 20-minute drone flights. We first show that our fast frame alignment method and change-aware object detection model substantially improve the speed-accuracy curve on three video sensing tasks: real-time parking availability mapping, pedestrian activity mapping, and street hazard detection. We then show that SkyQuery's cyclic architecture, where insights extracted from video are used to inform future routing decisions, is effective at achieving three diverse objectives on a parking availability mapping task. Finally, we demonstrate that SkyQuery is effective for a wide range of applications by showing qualitative results from three case studies. We release the SkyQuery code and datasets at \url{https://favyen.com/skyquery/}.

In summary, our contributions are:
\begin{itemize}[nosep]
\item A computer vision pipeline for converting drone video data into a symbolic representation suitable for querying (Section~\ref{sec:cv}).
\item A domain-specific language for expressing queries over a spatial region, along with an engine that executes these queries by flying drones to collect the necessary video data and running the query on this data (Section~\ref{sec:lang}).
\item An evaluation of our tool, SkyQuery, on three case studies, demonstrating how it can be used to express and execute interesting and useful queries (Section~\ref{sec:eval}).
\end{itemize}

\section{Overview}

\begin{figure}[t]
\begin{center}
	\includegraphics[width=\linewidth]{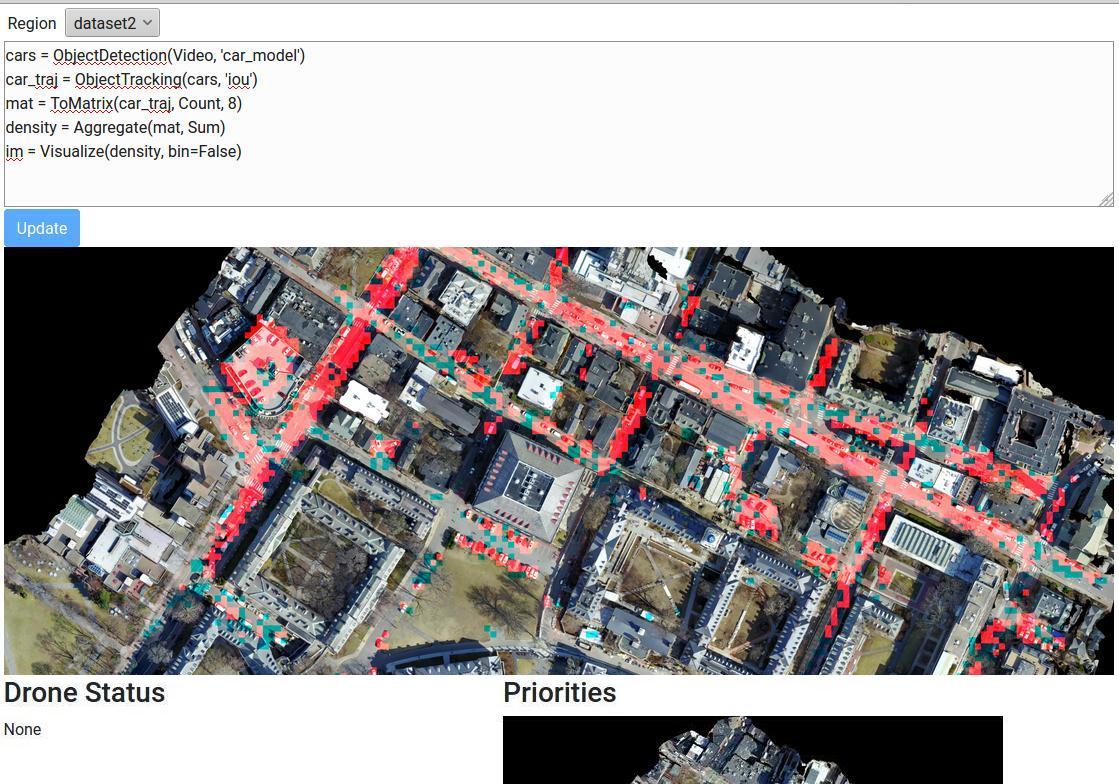}
\end{center}
	\caption{SkyQuery system interface.}
\label{fig:ui}
\end{figure}

In this section, we provide an overview of SkyQuery. To perform an aerial drone video sensing task, users first specify the geographic region that they are interested in observing. SkyQuery will then route drones to fly around the region and capture an initial stream of video. Once some video is available, users develop an analytics program using SkyQuery's interactive system interface. We show an example in Figure \ref{fig:ui}, where a user begins exploring their video by visualizing the density of detected cars across the region.

SkyQuery programs define sensing-analytics-routing loops, where video captured by aerial drones is analyzed to extract application-specific insights, and then further processed to decide where drones should fly and capture video next. Programs specify video analytics steps as a series of data processing operations that transform video into structured insights, represented as tabular streams that we call dataframes. Some of these dataframe streams are exported for downstream application processing or for visualization in the system interface. A special ``priorities'' dataframe is used for routing drones: it assigns a priority score to each cell in the region, and drones are routed so that high-scoring cells are visited more frequently.

Once a program is submitted for execution, the SkyQuery server computes cell priorities from the initial video sample, and begins assigning routes to idle drones based on those priorities. SkyQuery's drone controller runs on each drone to handle automatic takeoff, waypoint following, landing, and recharging. After a drone lands, it uploads captured video to the SkyQuery server over Wi-Fi, and SkyQuery processes the video to emit new rows to the output dataframes, update the priorities used to compute future routes, and re-render any active visualizations in the system interface.

\begin{figure}[t]
\begin{center}
	\includegraphics[width=0.85\linewidth]{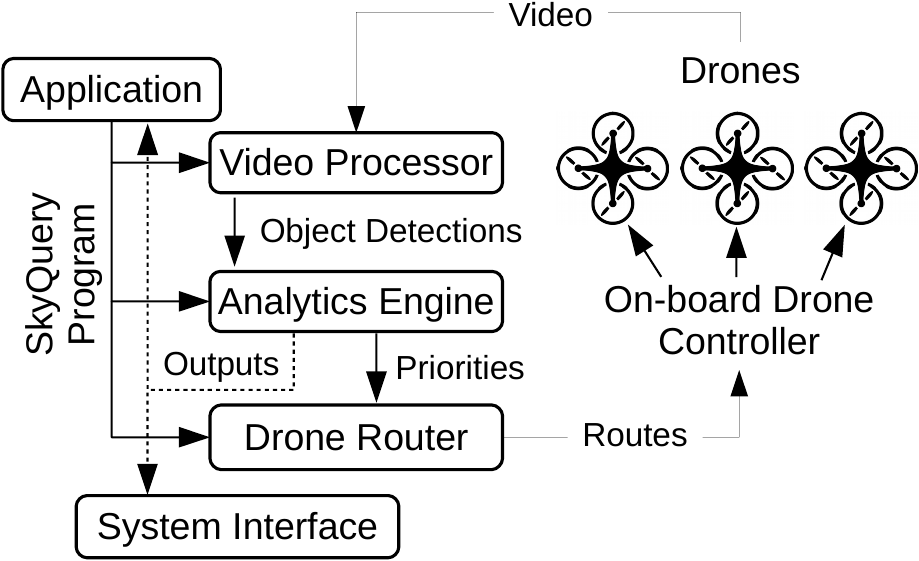}
\end{center}
	\caption{SkyQuery system architecture.}
\label{fig:arch}
\end{figure}

We summarize SkyQuery's system architecture in Figure \ref{fig:arch}. On the SkyQuery server, the video processor ingests video, and applies our novel fast frame alignment method and change-aware object detection model to extract structured information from raw video. This data is then passed to the analytics engine, which executes the application program and produces an output stream. This stream may include routing priorities, which SkyQuery accounts for when computing routes for idle drones.

\section{Related Work}

Several frameworks have been proposed to simplify the programming of aerial drones for sensing tasks~\cite{bregu2016, dedousis2018framework, tecola, voltron, buzz}.
Voltron~\cite{voltron} parallelizes actions that need to be performed independently at several locations across multiple drones, while satisfying application-specific time constraints.
Dolphin~\cite{dolphin} allocates tasks on subsets of drones, supporting rich task dependencies such as concurrent, sequential, and event-based execution.
Unlike SkyQuery, however, these frameworks focus on resource allocation and group-level control, and require users to implement complex sensor processing steps and route prioritization in the application code.

Several systems have tackled the problem of executing queries over video data~\cite{noscope, lu2018accelerating}. For example, BlazeIt~\cite{blazeit} implements a query language allowing users to select video frames through predicates that reason about the objects detected in the frame. However, prior work is limited to queries over object detections and tracks (e.g., ``show frames with exactly two cyclists and one car''). In contrast, the SkyQuery analytics engine implements a powerful query language that enables substantially more complex queries, including aggregation over time with spatial locality  and association of detections across different flights over a point.

Recent works have applied aerial drone video sensing for diverse applications, including identifying mosquito breeding sites~\cite{mosquito}, monitoring traffic conditions~\cite{traffic}, and detecting wildlife~\cite{wildlife}. These works highlight the need for a general-purpose aerial drone video sensing platform that simplifies the development of sensing applications. SkyQuery provides exactly such a platform.

\section{Video Processor}
\label{sec:cv}

The video processor runs on the SkyQuery server, and is responsible for extracting initial structured information from video, such as dataframes of object detections. To do so, it aligns video frames by computing a mapping from each pixel to world coordinates (longitude-latitude), and applies machine learning models to extract information from video.

\subsection{Fast, Accurate Frame Alignment} \label{sec:alignment}

Prior work in video frame alignment does not offer sufficient processing speed and accuracy for aerial drone video sensing applications.
Broadly, prior work can be classified into two categories. Offline mapping methods first compute a 2D or 3D map of the region, and then match video frames to this map~\cite{ibrahim2010moving}.
However, both computing the orthoimage and matching frames are slow. Other techniques apply simultaneous localization and mapping methods to localize the drone without a pre-existing map~\cite{tang2019gcnv2}. However, although these methods are ideal in GPS-denied settings, they offer subpar accuracy in the outdoor environments that we consider in our work because they do not leverage GPS and compass readings that are readily available in those environments.

In contrast, our fast, high-accuracy frame alignment method efficiently combines GPS and compass readings with image features without materializing an expensive orthoimage. We first summarize our algorithm, and then detail the steps below. We begin by applying SIFT feature extraction to derive keypoints in each video frame. A keypoint consists of a pixel position and a feature descriptor that describes a patch of the image around that position. We then group together keypoints that likely correspond to the same world location (e.g., a road marker or building chimney) in two phases. First, we create ``stable groups'' by grouping together keypoints of the same location in individual segments of video where the drone flies over the location.
Second, we create ``global groups'' by merging stable groups that describe the same world location, so that if a drone flies over a location several times, all of the corresponding keypoints are associated together.
Then, to align a video frame, we match SIFT keypoints computed in the frame against global groups, and use this matching to estimate the drone's position and orientation when it captured the frame.

\smallskip
\noindent
\textbf{SIFT Feature Extraction.} We apply SIFT feature extraction to compute keypoints in each video frame. A keypoint is a tuple $(p, f)$ consisting of a pixel position $p$ and a SIFT descriptor $f$ that describes the image in a patch around $p$. For every video frame $I_j$, SIFT feature extraction yields a set of keypoints $KP_j = \{(p_1, f_1), \ldots, (p_n, f_n)\}$.

\smallskip
\noindent
\textbf{Aggregating Keypoints into Stable Groups.} When a drone flies over a given world location, the drone captures several observations (keypoints) of that location in a subsequence of video frames. Because the drone may be moving, the location appears at a different pixel position in each video frame in this subsequence. To further process the SIFT keypoints, we must aggregate keypoints that likely correspond to the same world locations.

We develop an iterative algorithm to perform the aggregation. We maintain a set of stable groups $S = \{s_1, \ldots, s_m\}$ (initially empty), where each group $s$ is specified by the sequence of keypoints associated with the group. We iterate over the frames: for each frame $I_j$, we attempt to match each keypoint $(p, f) \in KP_j$ with an existing group $s$. We enforce two conditions when matching a keypoint to a group. First, we require $|f-s.f| < T_f$, where $s.f$ is the group's average feature descriptor computed by taking the mean across descriptors in the group, and $T_f$ is a threshold. Thus, the new keypoint's features should be similar to the features of keypoints that are already members of $s$. Second, if the most recent keypoint in $s$ appears at the pixel $p_g$ in $I_k$ ($k < j$), then we compute the optical flow $\Delta p$ from $I_k$ to $I_j$ at $p_g$ through the Lucas-Kanade method~\cite{lkflow}.
We require $d(p_g + \Delta p, p) < T_d$, where $d$ computes Euclidean distance between two pixels, and $T_d$ is a threshold.
Then, if $(p, f)$ matches to some existing group $s_i \in G$ under these conditions, we add $(p, f)$ to $s_i$. Otherwise, we create a new group $s = \langle (p, f) \rangle$, and add $s$ to $G$.

\smallskip
\noindent
\textbf{Estimating Spatial Coordinates of Stable Groups.} After aggregation, we use GPS and compass sensor readings to estimate the world coordinates of each group, i.e., its longitude, latitude, and height.
When computing world coordinates for a group $s$, our input includes both the keypoints $(p_i, f_i) \in s$ and the corresponding sensor readings, which include the spherical GPS coordinates transformed to a local reference frame, $(d_i.x, d_i.y, d_i.h)$ (2D Euclidean position in the local reference frame, and height), and the orientation $d_i.\alpha$ (measured counterclockwise from east).

We compute the camera angle $(\theta_{i,x}, \theta_{i,y})$ corresponding to each keypoint based on its pixel position and the camera parameters. Specifically, if a keypoint is extracted at $p_i = (x_i, y_i)$ (measured in pixels from the center of the image), the image resolution is $W \times H$, and the camera's horizontal field of view is $fov_x$, then the horizontal angle between the camera's normal axis (which corresponds to the world vertical axis, since we assume the camera points straight down) and the keypoint is $\theta_{i,x} = \arctan \frac{x_i \tan(fov_x/2)}{W/2}$.
We now estimate the group's world coordinates $(s.x, s.y, s.h)$. Each keypoint provides two linear equations relating these coordinates to the drone's coordinates:
\setlength{\abovedisplayskip}{3pt}
\setlength{\belowdisplayskip}{3pt}
\begin{align}
(d_i.h-s.h) \tan \theta_{i,x} & = s.x-d_i.x \\
(d_i.h-s.h) \tan \theta_{i,y} & = s.y-d_i.y
\end{align}
Then, we compute the three unknowns $(s.x, s.y, s.h)$ by solving using the least squares estimate.

\smallskip
\noindent
\textbf{Global Groups.} The sensor readings $(d_i.x, d_i.y, d_i.h)$ are inexact, and the estimates of stable group coordinates computed above may contain errors. To address this, we aggregate stable groups that correspond to the same world location into global groups. Then, we estimate the coordinates of a global group as the average across coordinates of the group members (stable groups). By averaging the coordinates over all of the instances where a drone flies over the location, we substantially reduce the error.

To merge stable groups into global groups, we repeat our aggregation algorithm above, but replace the optical flow constraint with a position estimate similarity constraint; given a stable group $s$ and a global group $g$:
$$\sqrt{(s.x-g.x)^2+(s.y-g.y)^2+(s.h-g.h)^2} < T_d$$
Here, $(g.x, g.y, g.h)$ are the mean coordinates of the global group, and $T_d$ is a threshold.

\subsection{Change-Aware Object Detection} \label{sec:detect}

\begin{figure}[t]
\begin{center}
	\includegraphics[width=\linewidth]{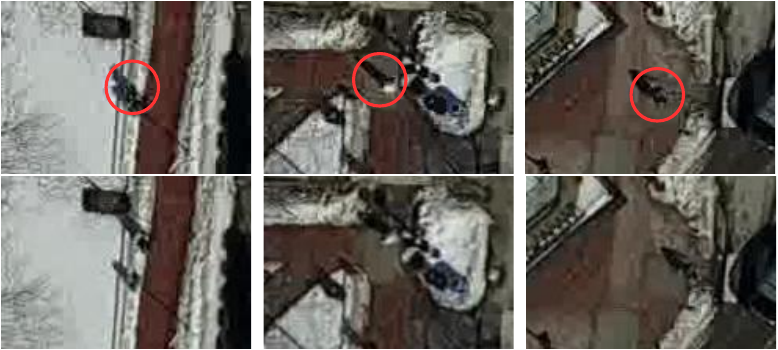}
\end{center}
	\caption{Example pedestrians in 1080p aerial drone video. Pedestrians are difficult to identify in still images even for humans, but become easier to identify by comparing multiple distinct frames (top and bottom images) close in time.}
\label{fig:ped}
\end{figure}

Object detection in aerial drone video is especially challenging because oftentimes, objects of interest, such as pedestrians and cyclists, have a small footprint in the top-down image. General-purpose object detection models such as YOLOv3 and Mask R-CNN offer state-of-the-art performance on detecting large and medium-sized objects that have distinctive appearances, but we show in Section \ref{sec:eval} that they yield poor performance on detecting pedestrians in aerial drone video: oftentimes, as we show in Figure \ref{fig:ped}, it is difficult even for humans to distinguish pedestrians.

To address this challenge, we develop a specialized model architecture, which we call change-aware object detection, for detecting small moving objects in aerial drone video. Our specialized model is motivated by two key insights.

First, we find that many potential false positive detections made by general-purpose models actually correspond to non-moving objects like street lights and fire hydrants. Thus, we can improve the robustness of object detection by comparing views of a location from different timestamps (Figure \ref{fig:ped}); in other words, through the training procedure, we enable our model to learn to become aware of change. Rather than input a single image, our detection model inputs three images: the current frame $I_j$, a recent frame from a small amount of time in the past $I_{j-\Delta t}$, and an earlier frame $I_k$ with $k << j$. $I_{j-\Delta t}$ provides short-term change information, enabling the model to identify corresponding image patches and analyze whether the motion of those patches is similar to the motion of the desired object type. $I_k$ provides long-term information, which helps both to improve the model's ability to eliminate static objects, and to enable the detection of objects like pedestrians and cyclists even when they are temporarily stopped.
We transform $I_{j-\Delta t}$ and $I_k$ using the drone positions estimated in Section \ref{sec:alignment} so that they align at surface height with $I_j$.

\begin{figure}[t]
\begin{center}
	\includegraphics[width=0.7\linewidth]{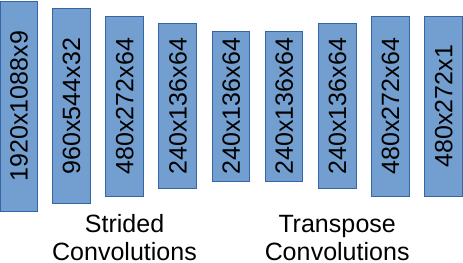}
\end{center}
	\caption{Simple, shallow neural network architecture we use for our change-aware object detection method.}
\label{fig:detect_model}
\end{figure}

Second, we use a simplified model architecture (Figure \ref{fig:detect_model}) that reduces the number of parameters that the model must learn. Standard object detection models such as YOLOv3 and Mask R-CNN output five predictions in each cell: estimates of the bounding box coordinates $t_x, t_y, t_w, t_h$, and an objectness confidence score $t_o$ indicating the likelihood that the center of an object falls in the call. Because object instances of types such as pedestrians and cyclists essentially all have the same width and height in aerial drone video, our model outputs only the center of the bounding box. Additionally, because the object sizes are small, we expect that the model is able to precisely localize the center of the object. Thus, rather than estimate the position $(t_x, t_y)$ within the cell, we simply use a small cell size, and only output $t_o$.

\section{Analytics Engine}
\label{sec:lang}

The analytics engine provides a domain-specific language for expressing aerial drone sensing tasks, and executes these programs over the initial dataframes extracted by the video processor. We show several example programs in Figure \ref{fig:programs}. Each line in a program produces a dataframe (left) by applying an operator (capitalized) on either video or one or more other dataframes. Programs specify an output dataframe that users can visualize in the SkyQuery interface or export as an output stream for usage in downstream applications.

Programs may also specify a priorities dataframe that defines how different areas of a region should be prioritized for future drone observation. SkyQuery uses these priorities to assign drone routes, so that drones fly more frequently over areas with high priority.

\begin{figure}[t]
    \begin{subfigure}{0.47\textwidth}
        \begin{flushleft}
            \mytext{\pt{cars} = ObjectDetection(Video, `car\_model')} \\
            \mytext{\pt{car\_traj} = ObjectTracking(\pt{cars})} \\
            \mytext{\pt{stopped} = Select(\pt{car\_traj}, displacement < 3)} \\
            \mytext{\pt{parked} = Select(Merge(\pt{stopped}), duration > 120)} \\
            \mytext{\pt{raw} = ToMatrix(\pt{parked}, Count, 64)} \\
            \mytext{\pt{spots\_bin} = Aggregate(\pt{raw}, Sum) > 0} \\
            \mytext{\pt{spots} = Thin(spots\_bin)} \\
            \mytext{\pt{rates}, \pt{counts} = ForecastRates(\pt{raw}, SimpleGaussian)}
            \mytext{\pt{priorities} = Aggregate(\pt{rates}, Priority)}
        \end{flushleft}
        \caption{Parking monitoring application. Produces \pt{spots} matrix of inferred parking spot locations, and \pt{counts} table of parked car counts.}

        \vspace{8pt}
        
        \begin{flushleft}
            \mytext{\pt{ped} = ObjectDetection(Video, `ped\_model')} \\
            \mytext{\pt{good} = Select(ObjectTracking(\pt{ped}), length > 3)} \\
            \mytext{\pt{new} = ToMatrix(\pt{good}, func=CountNew)} \\
            \mytext{\pt{activity} = Aggregate(\pt{new}, func=SUM)} \\
            \mytext{\pt{roads} = Thin(ToMatrix(\pt{car\_traj}, func=CountSum))} \\
            \mytext{\pt{crosswalks} = Import(`crosswalks.png')} \\
            \mytext{\pt{noncrosswalk} = \pt{roads} * (1 - \pt{crosswalks})} \\
            \mytext{\pt{ped\_road} = Join(\pt{good}, \pt{noncrosswalk})}
        \end{flushleft}
        \caption{Pedestrian activity mapping application. \pt{activity} is a matrix of pedestrian counts that SkyQuery visualizes as an activity map, and \pt{ped\_road} contains pedestrian sequences that cross roads outside of crosswalks, which can help inform crosswalk installation decisions.}
        
        \vspace{8pt}
        
        \begin{flushleft}
            \mytext{\pt{cycling\_lanes} = Import(`cycling-lanes.png')} \\
            \mytext{\pt{stopped} = Select(\pt{car\_traj}, displacement < 3)} \\
            \mytext{\pt{stopped} = Select(\pt{stopped}, duration > 5)} \\
            \mytext{\pt{hazards} = Join(\pt{stopped}, \pt{cycling\_lanes})}
        \end{flushleft}
        \caption{Street hazard detection application. \pt{hazards} contains sequences of cars that are stopped in cycling lanes.}
    \end{subfigure}
    
    \caption{Example SkyQuery programs. Tables are highlighted in magenta, and operators are capitalized.}
    \label{fig:programs}
\end{figure}

\subsection{Dataframe Types} \label{sec:types}

Each dataframe is a stream of rows that follow a schema based on the dataframe type. SkyQuery incorporates three dataframe types: detections, sequences, and matrices.

\smallskip
\noindent
\textbf{Detections.} This data type represents the results of applying an object detection method on video frames captured by a drone to obtain detections of a particular object type. Each row $(d.time,d.bounds)$ in the dataframe is a single detected object, where $d.time$ is the video frame timestamp and $d.bounds$ is the detection bounding box.

\smallskip
\noindent
\textbf{Sequences.} This data type represents sequences of object detections. Each row in a sequence dataframe is an ordered list $\langle d_1, \ldots, d_n \rangle$, where $d_i$ references a detection. Sequences capture relationships between detections over time. For example, the \texttt{ObjectTracking} operator links detections from consecutive frames that are close to each other, and thus are likely to be the same object.

\smallskip
\noindent
\textbf{Matrices.} This data type represents a spatio-temporal matrix, which divides the region into a grid with a user-specified cell size, and associates a time series to each cell. Each row $(o.cell, o.time, o.value)$ corresponds to one observation, where $o.cell$ is the observed cell, $o.time$ is the timestamp of the observation, and $o.value$ is the observed value for the given cell at the given time.

\subsection{Operators} \label{sec:operators}

We detail the key SkyQuery operators below. After each operator's name, we include it's input and output data types: we denote video $V$, detection dataframes $D$, sequence dataframes $S$, and matrix dataframes $M$.

\smallskip
\noindent
\textbf{ObjectDetection ($V \rightarrow D$).} This operator represents the outputs of an object detection model on aerial drone video. In the analytics engine, we implement ObjectDetection simply by pulling the corresponding data from the video processor. Rows in the result of this operator specify bounding boxes that are already transformed to world coordinates.

\smallskip
\noindent
\textbf{ObjectTracking ($D \rightarrow S$).} This operator applies an object tracking algorithm (specifically, SORT~\cite{sort}) to link detections of the same object instance across different frames of a video. Thus, ObjectTracking transforms detections into sequences by constructing tracks of multiple detections. For example, if a drone observes a car driving forwards along a road, ObjectDetection will yield a distinct detection for each frame containing the car, but ObjectTracking combines these detections into a single sequence. Object tracking operates locally, and if an object leaves the drone's camera frame and later returns, two distinct sequences will be constructed corresponding to each contiguous interval during which the object was visible.

\smallskip
\noindent
\textbf{Select ($S \rightarrow S$).} This operator selects sequences satisfying a user-specified criterion. For example, \texttt{length > 5} selects sequences with more than 5 detections, \texttt{displacement < 10} selects sequences with small displacement, and \texttt{duration > 60} selects sequences that span longer than 60 seconds. Select is frequently applied to filter sequences for ones that are of interest to an application. In Figure \ref{fig:programs}, we use Select in parking availability monitoring to select stopped cars and cars that are parked for at least two minutes, and in pedestrian activity mapping to prune pedestrians that were only detected in one or two frames.

\smallskip
\noindent
\textbf{Merge ($S \rightarrow S$).} This operator merges sequences that correspond to the same stationary object to produce a smaller set of longer sequences: if a drone flies over a parked car twice, ObjectTracking would yield a separate sequence during each flight, and Merge would merge them into one.

We implement Merge as an iterative algorithm. Let $S$ be the set of output (merged) sequences. Given a new sequence $s_2$ from the operand dataframe, we will specify a set of conditions that must be satisfied to merge $s_2$ into an existing sequence $s_1 \in S$.
Let $d_1$ be the last detection in $s_1$, and $d_2$ be the first detection in $s_2$. First, if the detections $d_1$ and $d_2$ do not overlap, then we do not merge $s_2$ into $s_1$. Similarly, if $d_2$ precedes $d_1$, we do not merge. These simple spatial and temporal constraints ensure that $s_2$ is in the same location as $s_1$, and at a later time.
However, these constraints are not sufficient to conclude that $s_1$ and $s_2$ correspond to the same object. For example, if $s_1$ and $s_2$ are parked cars, then $s_1$ may have left the parking spot, and $s_2$ is a different car. In fact, the simple constraints above would merge all cars that ever parked in the same spot together into one sequence.

To address this issue, we provide two additional constraints.
First, we do not merge if $d_1.bounds$ was visible in the drone's camera frame in some image captured after $d_1.time$ but before $d_2.time$. If we fly over a detected stationary object, fly over a second time and observe that the object is no longer there, and then fly over a third time and observe another stationary object, then we know that the two stationary objects are likely distinct.
Second, we compute an image similarity of the first and second detections (using structural similarity index~\cite{structsimindex}), and only merge $s_1$ and $s_2$ if the similarity score is above a threshold.

\smallskip
\noindent
\textbf{ToMatrix ($S \rightarrow M$).} This operator transforms a sequence dataframe into a matrix. Each matrix divides the region into a grid of cells under a user-configurable cell size. Then, given the set $R_{c,t}$ of sequences in the operand dataframe that intersect a cell $c$ at time $t$, we apply an aggregation function $g(c, t, R_{c,t})$ to compute the matrix value at grid cell $c$ and timestamp $t$. We use the aggregator to create a matrix observation $o$, where $o.cell = c$, $o.time = t$, and $o.value = g(c, t, S)$.

We provide several aggregators that the user can select. For example, the \texttt{Count} aggregator simply returns the number $|R_{c,t}|$ of sequences at $(c, t)$. Using \texttt{Count}, \texttt{ToMatrix} transforms the operand table of sequences into a matrix containing counts of those sequences (e.g., the number of pedestrians observed in each grid cell over time).

\smallskip
\noindent
\textbf{Join ($(S,M) \rightarrow S$).} This operator accepts two operands, a sequence dataframe and a matrix. It applies a spatio-temporal join between the sequences and the matrix, outputting only the sequences whose detections are contained in cells and at times for which the matrix is non-zero, i.e., sequences that intersect the matrix.

\smallskip
\noindent
\textbf{Aggregate ($M \rightarrow M$).} This operator applies an aggregation function on the time-varying values in each cell of a matrix, producing a new matrix containing the aggregated values. Like \texttt{ToMatrix}, \texttt{Aggregate} expects the user to provide an aggregation function parameter $g(O)$, which computes a value from a set of matrix observations $O$.

\subsection{Computing Priorities} \label{sec:priorities}

As new video is collected, SkyQuery leverages a user-defined \emph{priority dataframe} to adjust drone routes and prioritize drone flights in areas with more relevant or variable data. The priority dataframe is a matrix that assigns a priority to each grid cell; the priority of a cell should quantify the application's current preference for visiting that cell.

To simplify the assignment of priorities during application development, we provide a library that includes operators covering common use cases. These operators compute \emph{priority rates} that specify the rate at which the priority of a cell should increase. These rates are represented in a matrix dataframe, and correspond to the frequency that we want drones to fly to each grid cell. We also provide a \texttt{Priority} aggregation function for the \texttt{Aggregate} operator to compute priorities from priority rates, by adding up the priority rates over time, but resetting the priority at a cell to 0 when a drone observes the cell. We detail the rate operators below.

\smallskip
\noindent
\textbf{ConstRates ($\rightarrow M$).} The constant rates operator offers the simplest prioritization scheme: it sets the priority rate at all cells to 1, so that drone flights visit all cells in the region with equal frequency.

\smallskip
\noindent
\textbf{TTLRates ($M \rightarrow M$).} The time-to-live (TTL) operator ensures that drones only fly over cells that contain instances of an object type of interest. It accepts a user-defined parameter \texttt{ttl}, and expects a matrix operand $B$. Initially, \texttt{TTLRates} sets the rate at all cells to 1. However, after receiving \texttt{ttl} observations in $B$, such that each observation $o$ has $o.value = 0$, we set the rate at that cell to 0 in the future so that drones stop visiting the cell. On the other hand, if we receive an observation $o$ with $o.value > 0$, then \texttt{TTLRates} will set priority rates so that drones continue to visit $o.cell$ indefinitely.

This operator is useful in cases where the user wants drones to only fly over areas with activity relevant to the application, but does not know a priori where these areas are. For example, when monitoring parking spots, we want to minimize the time that drones spend flying in areas with no parking spots. Thus, we could compute a matrix of parked car counts, and then provide this matrix to \texttt{TTLRates} so that cells with no parked cars are de-prioritized.

\smallskip
\noindent
\textbf{ForecastRates ($M \rightarrow (M,M))$.} The forecasting operator prioritizes drone flights in cells where values in the operand matrix are less predictable. Continuing the parking monitoring example, the user may find that substantial drone flight time is wasted flying frequently over residential and office parking lots where parking follows predictable patterns --- an office lot may have a predictable spike in parking after the morning rush hour, and a predictable drop after the work day ends. Thus, the user may want to instead prioritize cells with less predictable parking patterns, such as retail lots, where e.g. parking may increase due to a sale event.

Thus, we apply a forecasting model (such as Gaussian process model or ARIMA) to predict future values in the operand matrix $B$. We treat the values of observations in a particular cell $c$ as a time series that is sparsely sampled. We then fit the forecasting model to predict changes in the time series values. If the set of observations at $c$ is $O_c = \langle o_1, \ldots, o_n \rangle$, then we fit the model to a time series of differences $\langle o_2.value - o_1.value, \ldots, o_n.value - o_{n-1}.value \rangle$. We then apply the model to predict the current matrix value at all cells, even if drones are not currently observing the cell. Each prediction is a probability distribution. We set the priority rate to the variance of that distribution, which corresponds to the uncertainty in the model. Then, cells that exhibit unpredictable patterns in matrix values (e.g. parked car counts) will be assigned higher priority rates, and thus be visited more frequently.

In addition to outputting the priority rates computed based on forecasting, this operator also outputs a matrix containing the values predicted by the same forecasting model, which the user can opt to use in lieu of the original matrix for downstream processing.

\subsection{Routing Drones} \label{sec:routing}

We formulate the task of prioritizing high-priority cells during route assignment as an instance of the vehicle routing problem (VRP). We solve this problem to assign a route each time an idle drone is ready to take off. Specifically, in our VRP instance, visiting a cell provides a reward equal to the current priority of that cell, and the routing objective is to maximize the reward collected by the drone, with two constraints: the route must start and end at the drone depot, and total traveling time must be less than the battery life.
This VRP does not have a polynomial-time solution, but polynomial approximation algorithms are well studied in the vehicle routing literature~\cite{vehiclerouting}. We solve the VRP using the best-insertion algorithm implemented in Google OR-Tools~\cite{ortools}.

\subsection{Example Program}

We now detail the example parking monitoring program in Figure \ref{fig:programs}(a).
The program detects parking spots (in \texttt{spots}) by finding areas where cars stop for at least two minutes. It also outputs a matrix of parked car counts over time (\texttt{counts}).
As SkyQuery collects more video, it emits new observations of parked car counts and parking spots in the \texttt{counts} and \texttt{spots} dataframe streams.

The program first applies the \texttt{ObjectDetection} operator to produce a detection dataframe, \texttt{cars}, containing the bounding polygons of car objects detected in the video. Here, ``car\_model'' references an object detection model that the user has trained and registered with the SkyQuery platform; SkyQuery includes scripts to automatically train and register models using our change-aware detection approach on annotated images. We then apply \texttt{ObjectTracking} to associate detections of the same car across consecutive video frames, yielding the \texttt{car\_traj} sequence table of car trajectories. Since we are only interested in parked cars, we then apply \texttt{Select} to exclude sequences with large displacements (i.e., cars that moved).

However, since object tracking operates in local segments of video, a parked car observed in one flight will be represented in \texttt{stopped} by a separate sequence than the same car observed in a later flight. The \texttt{Merge} operator merges repeated sequences of the same stationary object. Then, we apply a second \texttt{Select} to exclude cars that are only briefly stopped, e.g., cars waiting at a stop light.

To detect parking spots, we compute a binary matrix \texttt{spots\_bin} where a cell has value 1 if it intersects a car sequence in \texttt{parked}. We apply thinning to derive a matrix \texttt{spots} where each parking spot corresponds to one cell.
To compute parked car counts, and route drones to fly more frequently over regions with unpredictable parking patterns, we apply the \texttt{ForecastRates} priority rate operator.

\section{System Setup}

\textbf{SkyQuery Interface.} The user interacts with SkyQuery through a web interface that runs on the SkyQuery server (Figure \ref{fig:ui}). From here, users can specify a program for their application, and experiment with different data processing workflows on previously captured video.

\begin{figure}[t]
\begin{center}
	\includegraphics[width=0.7\linewidth]{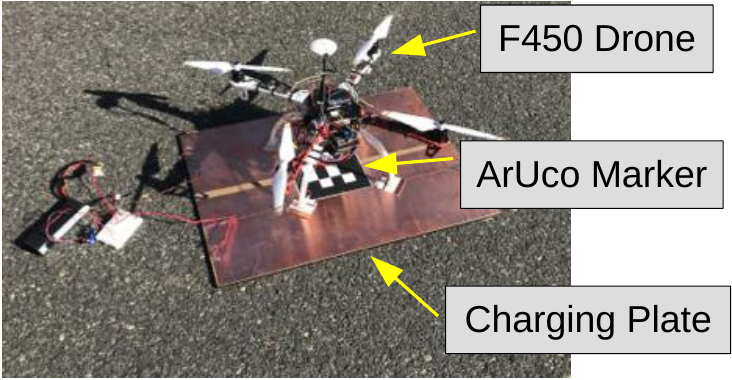}
\end{center}
	\caption{The SkyQuery aerial drone platform.}
\label{fig:drone}
\end{figure}

\smallskip
\noindent
\textbf{Aerial Drone Platform.} Our aerial drone platform (Figure \ref{fig:drone}) consists of two components to enable the autonomous operation of drones managed by SkyQuery: the drone itself, which is operated by our drone controller software, and the landing pad, which consists of a charging plate for re-charging drones and a WiFi router for uploading video captured by drones on their last flights. Although drone flight controllers natively support autonomous takeoff and landing, these controllers conduct landing based only on GPS and so their landing precision is not accurate enough to ensure that the drone does not collide with other drones and lands directly on its charging plate (to support autonomous recharging, we expose contact wires on the legs of the drones). Thus, our drone controller incorporates a precision landing procedure that leverages an ArUco optical marker~\cite{aruco} and PID controller.

\section{Evaluation} \label{sec:eval}

Users can develop programs on SkyQuery to perform a wide range of drone video sensing tasks. We evaluate the speed, accuracy, and efficacy of SkyQuery on three such tasks, implemented as SkyQuery programs in Figure \ref{fig:programs}: parking monitoring, pedestrian activity mapping, and street hazard detection.
To provide data for these applications, we capture aerial drone video over an urban area with variable car and pedestrian activity across eight 20-minute drone flights, and derive insights for each application from this video. We also hand-label 1,965 cars and 1,775 pedestrians in the video dataset to train our change-aware object detection method.
We will release the SkyQuery code and dataset.

We conduct four experiments to thoroughly evaluate each component of our system. In Section \ref{sec:eval_align}, we replace SkyQuery's frame alignment method (Section \ref{sec:alignment}) with each of three baselines and evaluate the impact on the speed-accuracy curve. In Section \ref{sec:eval_detect}, we compare the detection accuracy of our change-aware object detection method (Section \ref{sec:detect}) against two state-of-the-art general-purpose object detection methods. In Section \ref{sec:eval_schedule}, we conduct a simulation experiment involving a large-scale drone deployment using parking data recorded by the City of San Diego to evaluate the performance of SkyQuery's scheduling operators. Lastly, in Section \ref{sec:cases}, we demonstrate three case studies.

\subsection{Frame Alignment} \label{sec:eval_align}

\begin{figure}[t]
\begin{center}
	\includegraphics[width=\linewidth]{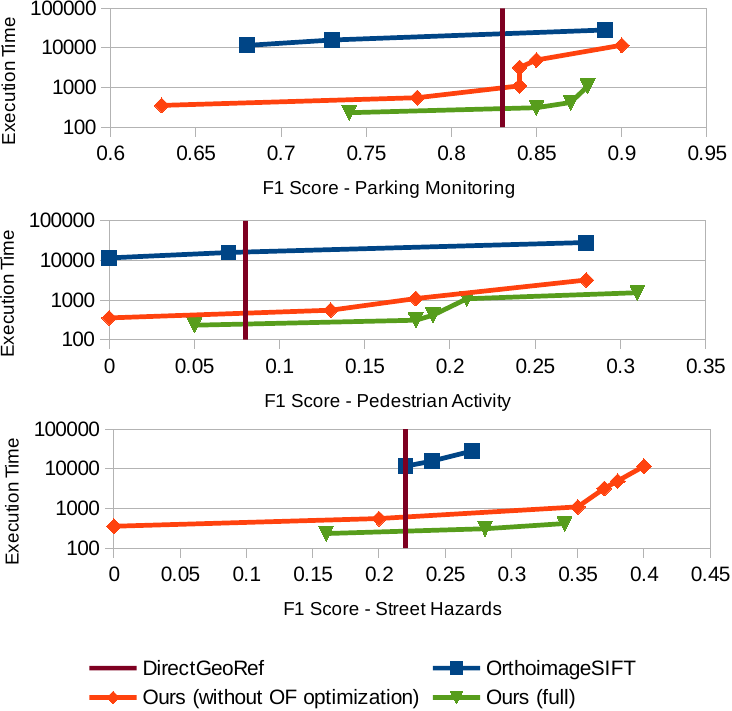}
\end{center}
	\caption{Curves showing execution time and accuracy provided by our method and the baselines on three applications. DirectGeoRef does not consider image cues in localizing the drone, and so executes virtually instantaneously; thus, we show its performance as a vertical line.}
\label{fig:eval_align}
\end{figure}

We first compare the performance of our fast frame alignment method to three baselines on each application. We evaluate methods in terms of their speed-accuracy tradeoff. To measure accuracy, we first compute end-to-end outputs through SkyQuery; here, we keep the SkyQuery components the same other than the alignment method (i.e., we use our change-aware object detection method and analytics engine). We then compare these outputs to ground truth:

\begin{itemize}[leftmargin=*,noitemsep,topsep=0pt]
    \setlength{\parskip}{0pt}
    \item Parking Monitoring: we apply SkyQuery to detect parking spots, represented by the binary matrix \pt{spots} in Figure \ref{fig:programs}(a), by identifying locations where cars remain stopped for two minutes. We compare \pt{spots} under an F1 score to the City of Cambridge parking spaces dataset~\cite{cambridge}, which includes the GPS locations of parking spots along two commercial corridors in the region where we captured video.
    \item Pedestrian Activity Mapping: we apply SkyQuery to detect pedestrian sequences that cross roads outside of crosswalks (\pt{ped\_road} in Figure \ref{fig:programs}(b)). We then hand-label the location and timestamp of each such pedestrian instance in the video data, and compare \pt{ped\_road} to the hand-labeled ground truth data in terms of F1 score.
    \item Street Hazard Detection: we apply SkyQuery to detect cars stopped in cycling lanes (\pt{hazards} in Figure \ref{fig:programs}(c)). As above, we hand-label these instances in the video dataset, and compare in terms of F1 score.
\end{itemize}

We compare against three baselines, where we replace our frame alignment method with the baseline procedure, but keep other components of SkyQuery the same. DirectGeoRef implements the direct geo-referencing approach in~\cite{eugster2007geo}, where the camera position is estimated directly from the position, altitude, attitude, and heading reported by the drone's GPS, IMU, and other sensors. OrthoimageSIFT implements the image alignment module in~\cite{ibrahim2010moving}, which applies SIFT feature matching to compute a geometric transformation from a captured video frame to real-world coordinates. ``Ours (without OF optimization)'' implements our method, but without the optical flow optimization introduced in Section \ref{sec:alignment}.

We show results for the three SkyQuery applications in Figure \ref{fig:eval_align}. We vary the image resolution at which each alignment method ingests the video frames to derive speed-accuracy curves. For OrthoimageSIFT, we consider only the SIFT feature matching time when processing video, and exclude the expensive orthoimage computation cost. DirectGeoRef yields poor accuracy due to GPS and compass noise.
Our method consistently outperforms OrthoimageSIFT: at the highest accuracy levels on the right of the charts, our method without the optical flow optimization provides a 3-4x improvement in execution speed, while leveraging optical flows provides an additional 3x speedup (10x total).

\subsection{Change-Aware Detection} \label{sec:eval_detect}

\begin{figure}[t]
\begin{center}
	\includegraphics[width=\linewidth]{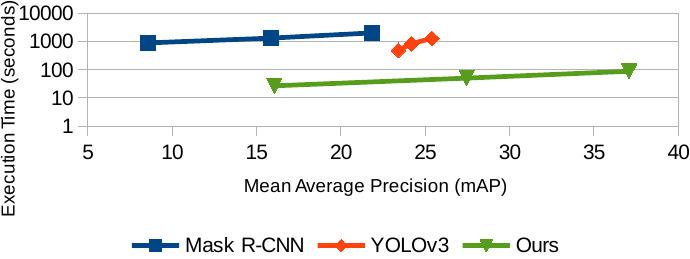}
\end{center}
	\caption{Mean average precision and execution time of detecting pedestrians in aerial drone video.}
\label{fig:eval_detect}
\end{figure}

We now evaluate our change-aware object detection method against YOLOv3~\cite{yolov3} and Mask R-CNN~\cite{maskrcnn} in terms of execution time and mean average precision (mAP), a standard detection metric that computes the average precision at varying recalls. We split our 1,775 hand-labeled pedestrian annotations into 80\%-10\%-10\% training, validation, and test sets. We train and test each object detector at several resolutions to derive a speed-accuracy curve. While we compute mAP on the test set, we measure execution time on inferring pedestrians over the entire three-hour video dataset, since the small size of the test set is insufficient for accurate timing measurements.

We show results in Figure \ref{fig:eval_detect}. Because our method targets objects of consistent sizes, it allows for employing a shallower neural network architecture, and so our method greatly outperforms the baselines in speed. This architecture is effective for detecting objects in aerial drone video because these objects generally do have consistent sizes.
Moreover, though, our change-aware method provides higher accuracy than the baselines at the full video resolution.

\subsection{Scheduling Operators} \label{sec:eval_schedule}

\begin{figure}[t]
\begin{center}
	\includegraphics[width=\linewidth]{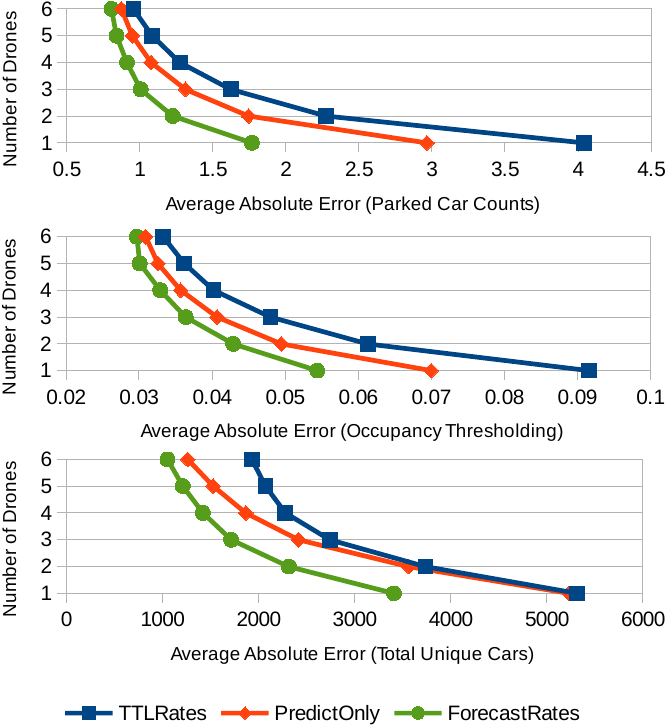}
\end{center}
	\caption{Average error on each of the three objectives.}
\label{fig:eval_schedule}
\end{figure}

\begin{figure*}[t]
    \begin{subfigure}{0.48\textwidth}
        \includegraphics[width=\linewidth]{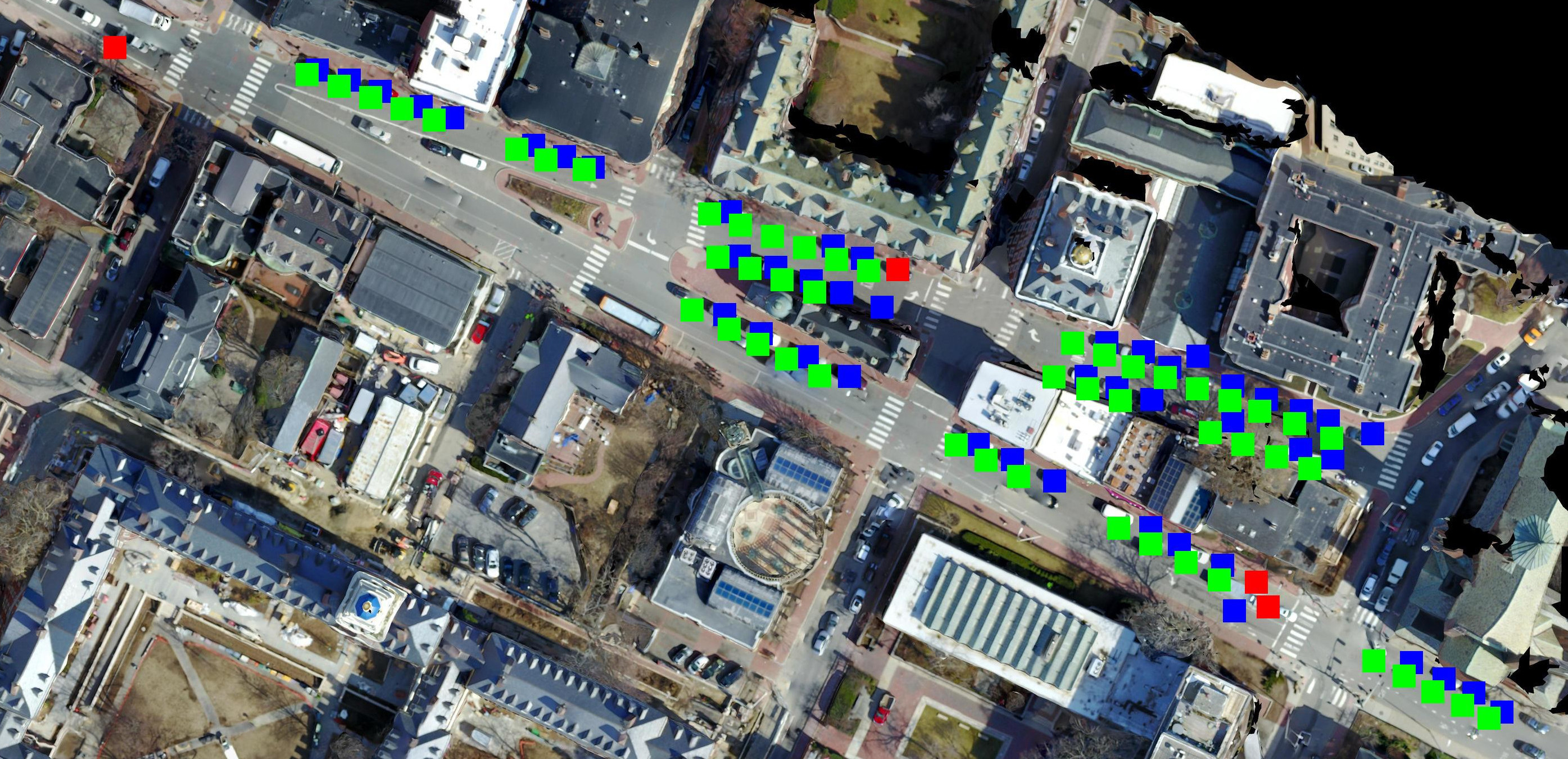}
        \caption{Comparison of the parking spots inferred by SkyQuery (\pt{spots}) with the city parking dataset. The city dataset is shown in green. Detected spots that match are colored blue, while incorrect detections are colored red.}
        
        \includegraphics[width=\linewidth]{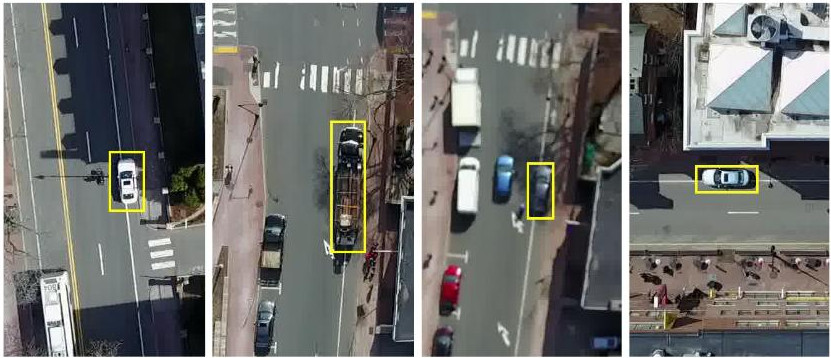}
        \caption{Sequences in \pt{hazards} showing cars stopped in cycling lanes.}
    \end{subfigure}
    \hspace{3mm}
    \begin{subfigure}{0.48\textwidth}
        \includegraphics[width=\linewidth]{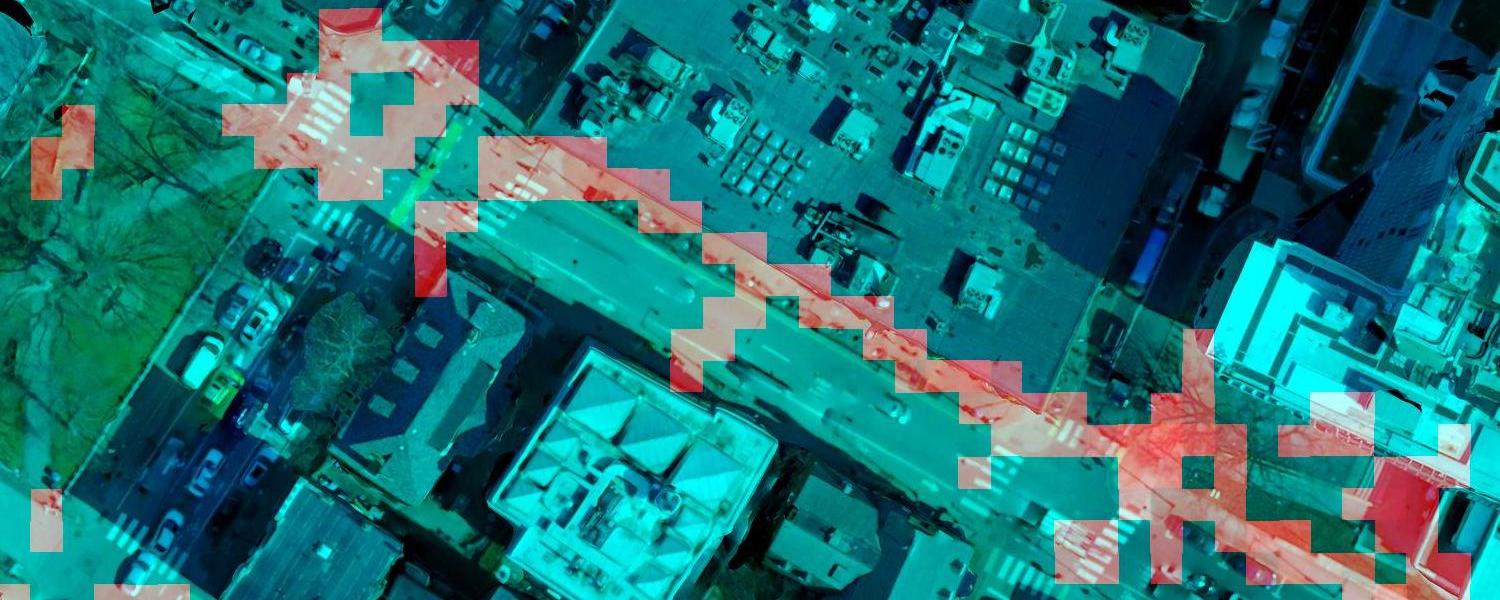}
        \includegraphics[width=\linewidth]{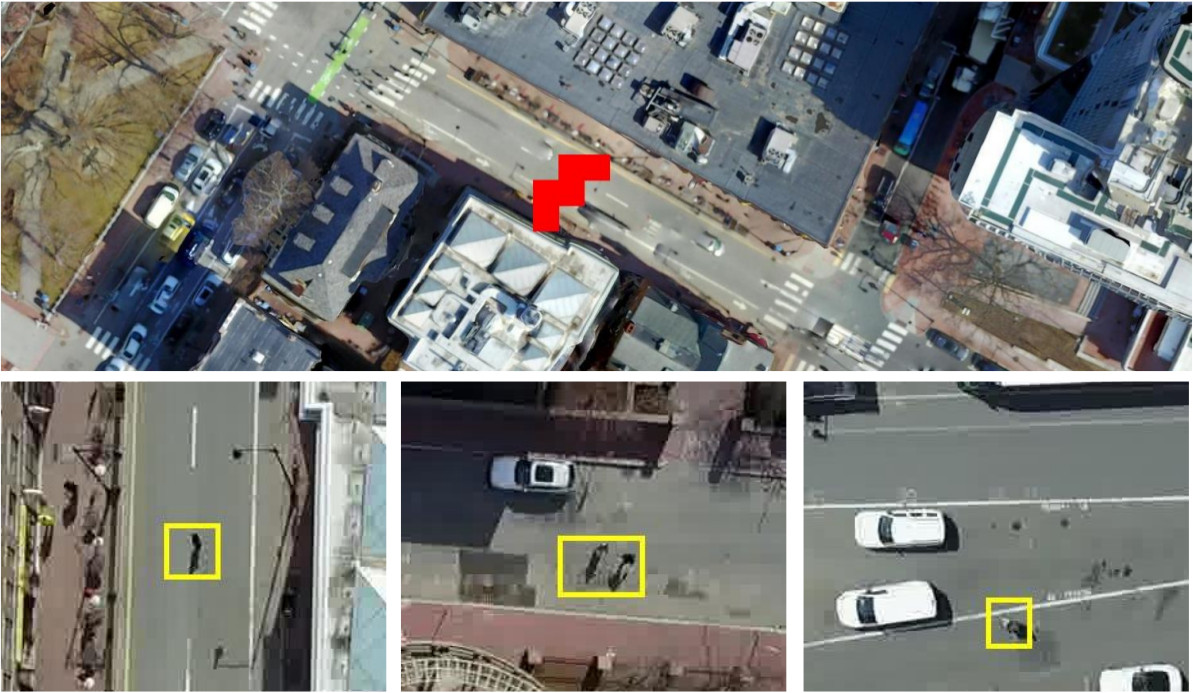}
        \caption{At top, SkyQuery's visualization of \pt{ped\_activity}, where cells with detected pedestrians are red. Below, visualizations of a matrix showing locations where we detect pedestrians crossing roads outside of crosswalks (middle), and sequences in \pt{ped\_road} associated with these detections (bottom).}
    \end{subfigure}
    
    \caption{Visualizations produced by the SkyQuery system of output tables from the example video sensing applications.}
    \label{fig:qual}
\end{figure*}

Because acquiring enough ground truth data and performing a sufficient number of flights to effectively evaluate the scheduling operators is too costly, we employ simulated flights. We build a simulator using an eight-week extract of metered parking data from the City of San Diego~\cite{sandiego}, which includes the start and end times and longitude-latitude of each parking event. We simulate between one and six drones flying across the city over the eight weeks. In the simulator, drones fly between cells of a 512 m $\times$ 512 m grid graph, traveling at 40 mph with a one-hour battery life and 256 m $\times$ 256 m field of view. Drones return to a depot in the center of the area covered by the dataset (which spans 150 $\text{km}^2$) to recharge.

To demonstrate the adaptability of the scheduling operators to different routing objectives, we consider three objectives related to parking monitoring:

\begin{itemize}[leftmargin=*,noitemsep,topsep=0pt]
    \setlength{\itemsep}{0pt}
    \item Parked car counts, where we monitor the number of parked cars in each region of the city.
    \item Regions with open spots, where we identify regions where at least half of the spots are available.
    \item Total cars, where we observe the total number of unique cars that have parked in each region.
\end{itemize}

Each objective involves accurately observing a matrix in the program below, which extends Figure \ref{fig:programs}(a); we mark the output matrices corresponding to each objective above with an asterisk:

\begin{myquery}
\mytext{\pt{parked} = Select(Merge(\pt{stopped}), duration>120)} \\
\mytext{*\textbf{\pt{counts}} = ToMatrix(\pt{parked}, Count)} \\
\mytext{\pt{capacities} = Aggregate(\pt{counts}, Max)} \\
\mytext{*\textbf{\pt{open}} = \pt{counts} > \pt{capacities}/2} \\
\mytext{\pt{new} = ToMatrix(\pt{parked}, CountNew)} \\
\mytext{*\textbf{\pt{total}} = Aggregate(\pt{new}, Sum)}
\end{myquery}

\noindent
When a drone flies over a cell in the simulation, it observes the current value of the \texttt{counts} matrix at that cell. We create three versions of this program corresponding to the three scheduling objectives by applying ForecastRates on each of \texttt{counts}, \texttt{open}, and \texttt{total}.
We evaluate each version independently, and measure performance on the corresponding objective in terms of absolute error between the outputs and the ground truth, averaged over space (the spatial regions) and time.

In Figure \ref{fig:eval_schedule}, we show results on the three objectives. We show the number of drones on the y-axis to highlight reductions in the number of drones needed to achieve particular error rates. The TTLRates baseline applies the time-to-live operator to assign routes to drones that visit all regions containing parking spaces with equal frequencies. PredictOnly leverages our forecasting method to predict intermediate time series values, but does not consider time series predictions when routing drones.
Our ForecastRates approach reduces the number of drones needed to obtain a particular error by 1.7$\times$ to 3$\times$ compared to the naive approach of visiting all regions with equal frequencies.
Thus, applications benefit substantially because SkyQuery reduces the number of drones that must be employed to manageable amounts.

\subsection{Case Studies} \label{sec:cases}

While our frame alignment, change-aware detection, and scheduling methods provide performance benefits for video sensing, our primary contribution is developing a system that makes it easy to both discover new insights in aerial drone video and express programs that extract those insights. Without our system, today, to experiment with different sensing tasks, users would need to manually align the video, experiment with and run object detectors, transform the detection coordinates, and develop custom visualization tools.

Thus, to demonstrate SkyQuery's overall effectiveness at enabling video sensing tasks, in Figure \ref{fig:qual}, we show qualitative outputs from the three video sensing applications. These outputs are visualizations produced by SkyQuery, but downstream applications may also ingest dataframe streams for further processing. In general, the data types and operations in SkyQuery succeed in enabling a wide range of aerial drone video sensing applications, from representing and computing the outputs of machine learning models (object detector outputs in detection dataframes, image classification outputs in matrix dataframes, etc.) to transforming those raw outputs into insights directly useful for applications, and prioritizing aerial drone flights spatially across a large region.

\section{Conclusion}

In this paper, we presented SkyQuery, a general-purpose platform for developing and deploying aerial drone sensing applications. SkyQuery programs define sensing-analytics-routing loops, where captured video is analyzed to derive application-specific insights, and these insights are further processed to decide where to route drones in the future. By transparently addressing the robust and efficient alignment of video and detection of objects, SkyQuery simplifies program development and promises to enable the exploration of a wide range of novel sensing tasks.

\end{sloppypar}

\bibliographystyle{ACM-Reference-Format}
\bibliography{main}

\end{document}